\documentclass[twocolumn,10pt,prl,showpacs]{revtex4-1}

\usepackage{amssymb, amsmath, amsthm}
\usepackage{graphicx}
\usepackage{subcaption}
\usepackage[all]{xy}
\usepackage{enumerate}
\usepackage[pdftex,hyperref,svgnames]{xcolor}
\usepackage[pdftex,colorlinks=true,
pdfstartview=FitV,
pdfnewwindow=true,
linktoc = page,
linkcolor= Red,
citecolor= blue,
urlcolor= blue,
hyperindex=true,
hyperfigures=false]{hyperref}
\hypersetup{linktocpage}
\usepackage{dsfont}
\usepackage{empheq}
\usepackage{float}
\usepackage{cancel}
\usepackage{relsize}
\usepackage{soul}
\usepackage{enumitem}
\usepackage{hhline}

\newcommand{\beq}{\begin{equation}}
\newcommand{\eeq}{\end{equation}}
\def\bea#1\eea{\begin{align}#1\end{align}}
\newcommand{\nn}{\nonumber}

\newcommand{\w}{\wedge}

\newcommand{\f}[2]{f^{#1}{}_{#2}}
\newcommand{\Rc}{\mathcal{R}}

\newcommand{\cgr}[1]{\textcolor{ForestGreen}{#1}}
\newcommand{\cre}[1]{\textcolor{red}{#1}}

\def\d {{\rm d}}
\def\mmm {\mathcal{M}}

\begin{document}

\title {\Large\bf Intricacies of classical de Sitter string backgrounds}
 \author{\bf David Andriot$^{a}$, Paul Marconnet$^{b}$, Timm Wrase$^{a,c}$}
\affiliation{$^{a}$Institute for Theoretical Physics, TU Wien,\\ Wiedner Hauptstrasse 8-10/136, A-1040 Vienna, Austria}
\affiliation{$^{b}$Ecole Normale Sup{\'e}rieure de Lyon,\\ 15 parvis Ren{\'e} Descartes, 69342 Lyon, France}
\affiliation{$^{c}$Department of Physics, Lehigh University,\\ 16 Memorial Drive East, Bethlehem, PA, 18018, USA}
\affiliation{\upshape\ttfamily david.andriot@tuwien.ac.at, paul.marconnet@ens-lyon.fr, timm.wrase@lehigh.edu}

\begin{abstract}
Up-to-date, there is no known example of a classical de Sitter solution of string theory, despite several good candidates. We consider here two newly discovered 10d supergravity de Sitter solutions, and analyse in great detail whether they can be promoted to classical string backgrounds. To that end, we identify five requirements to be met, and develop the necessary 10d tools to test the solutions. Eventually, they both fail to verify simultaneously all requirements, in spite of positive partial results. The explicit values obtained offer a clear illustration of the situation, and the analysis highlights various subtleties. We finally discuss the relation to the problem of scale separation.
\end{abstract}

\maketitle

\section{I. Introduction}

Obtaining a background of string theory with a four-dimensional (4d) de Sitter space-time in a well-controlled framework is challenging \cite{Danielsson:2018ztv}. This becomes an issue if one tries to reproduce from string theory -- as a candidate for a fundamental theory of nature -- standard models of cosmology, which describe the early or the future universe with (quasi) de Sitter space-times. This problem has been recently revived, not only by cosmological observations becoming more and more accurate, but also on the theory side with the swampland program. In the latter, conjectures were proposed asserting that any theory of quantum gravity, like string theory, and its 4d low energy effective theories, cannot admit solutions with a (quasi) de Sitter space-time \cite{Obied:2018sgi}, at least in some limit \cite{Ooguri:2018wrx, Bedroya:2019snp} that could correspond to a classical perturbative regime. In this paper we revisit this situation and claims in great detail on interesting examples. We focus on 10d supergravity solutions admitting a 4d de Sitter space-time, and we test to a rare extent all requirements for them to be a classical string background. Even though the solutions do not pass all tests, this detailed analysis should provide useful tools for such studies, as well as a better understanding of the mechanisms behind this apparent obstruction.

Finding classical de Sitter solutions of string theory is usually done in two steps. First, one typically looks for solutions of 10d type II supergravities with a 4d de Sitter space-time and a compact 6d manifold $\mmm$. Such solutions with intersecting $D_p$-branes and orientifold $O_p$-planes were found on $\mmm$ being a group manifold in \cite{Caviezel:2008tf, Flauger:2008ad, Caviezel:2009tu, Danielsson:2010bc, Danielsson:2011au, Roupec:2018mbn, Andriot:2020wpp}. The second step is to verify that these supergravity solutions are classical string backgrounds, by satisfying a list of requirements, including a small string coupling $g_s$ and a large 6d volume in units of the string length $l_s$. Studies on this topic were so far rather negative \cite{Roupec:2018mbn, Junghans:2018gdb, Banlaki:2018ayh, Andriot:2019wrs, Grimm:2019ixq} despite possible loopholes, leading to the situation that no stringy classical de Sitter solution has been found, in line with the swampland conjectures. In the companion work \cite{Andriot:2020wpp}, we have obtained new de Sitter solutions of 10d type IIB supergravity with intersecting $O_5/D_5$ sources, thus fulfilling the above first step. In this work, we provide a detailed 10d analysis of the second step, on two of these new solutions.

Group manifolds are characterised by an underlying Lie algebra with structure constants $f^a{}_{bc}$. The latter can be related to the spin connection coefficients of the 6d metric, as can be seen e.g.~in the Maurer-Cartan equations $\d e^a = -\frac{1}{2} f^a{}_{bc} e^b\w e^c$. The $f^a{}_{bc}$ then enter our 10d equations as variables. When looking for solutions in \cite{Andriot:2020wpp}, we allowed for a maximal freedom in the $f^a{}_{bc}$, in a basis of one-forms $e^a$ where the metric was $\delta_{ab}$. This simplified the search of solutions, but had the drawback of making the identification of the algebra and group manifold cumbersome. We typically got many non-zero structure constants, while Lie algebra representatives in classification tables only have a few. Algebras of some solutions could still be identified, with an isomorphism to our sets of $f^a{}_{bc}$. We could then verify the existence of a lattice, a discrete action on the 6d group manifold that provides its compactness. Algebras and lattices are particularly simple for the de Sitter solutions 14 and 15 of \cite{Andriot:2020wpp}, providing us, up to a change of basis, with a complete knowledge of their 6d geometry. So we focus here on solution 14, and treat solution 15 in the appendix. We briefly comment on algebras of other solutions at the end of the appendix.

To ensure that a 10d supergravity solution is a classical string background, we identify in practice five requirements to be met: a small $g_s$, large 6d radii $r^{a \, = \, 1, \,\dots, \, 6}$ in units of $l_s$, quantization of fluxes, a fixed number of orientifolds $N_{O_5}^I$, and lattice quantization conditions. We believe these requirements are sufficient; whether they are necessary is discussed at the end of section V. The last three requirements need the detailed 6d geometry, so they were only partially checked on the solutions in \cite{Andriot:2020wpp}. We now complete the study for solutions 14 and 15. We summarize the requirements as follows
\bea
& g_s \ll 1 \ , \ \ r^a \gg 1 \ ,\ \ N_{q\, a_1 \dots a_q} \in \mathbb{Z} \ ,\label{requir} \\
& N_s^I\in \mathbb{Z}\ ,\ \  N_s^I \leq N_{O_5}^I\ ,\ \ N_{a} \ \mbox{quantized} \ ,\nn
\eea
where for simplicity, we will choose a hierarchy factor of $10$, i.e.~$0 < g_s \leq 10^{-1} , r^a \geq 10 $. The index $I=1,2,3$ refers to the sets of internal dimensions $(ab)$ wrapped by the $O_5/D_5$ sources: $I=1$ along $(12)$, $I=2$ along $(34)$, $I=3$ along $(56)$. The number of sources along each set is defined as $N_s^I=N_{O_5}^I - N_{D_5}^I$. It enters the source contributions $T_{10}^I$ to the equations of motion and Bianchi identities. Our solutions have non-zero fluxes $F_1, F_3, H$, giving the flux integers $N_{q\, a_1 \dots a_q}$ along some of their components. Finally, the $N_a$ are numbers entering the $f^a{}_{bc}$, quantized because of the lattice. The relations to the supergravity solution data (in the left-hand side below) go as follows
\bea
& g_s F_{q\, a_1 \dots a_q}= \frac{g_s\, \lambda\, N_{q\, a_1 \dots a_q}}{r^{a_1} \dots r^{a_q}} \ , \label{qtties}\\
& g_s T_{10}^I = \frac{6\, g_s\, \lambda^2\, N_s^I}{r^{a_{1\bot_I}} \dots r^{a_{4\bot_I}}}\ ,\quad f^a{}_{bc} = \frac{r^a\, \lambda\, N_{a}}{r^b r^c} \ ,\nn
\eea
without sum on the indices, and the indices $a_{\bot_I}$ denote the directions transverse to the set $I$. The three types of supergravity variables (fluxes, source contributions, structure constants) are here expressed in units of $2\pi l_s$. The radii $r^a$ are introduced through a normalisation convention of the one-forms $e^a$ that we will come back to. The parameter $\lambda >0$ is an overall rescaling parameter of the solution that we are free to introduce. We refer to section 4 of \cite{Andriot:2020wpp} for more details \cite{foot0}.

In the following, we test solutions 14 and 15, expressed in the appropriate basis, upon the requirements \eqref{requir}. To that end, we provide in section II the material needed by first discussing the 6d geometry of the group manifold for solution 14. We give a basis of globally defined one-forms, determine lattice quantization conditions, and discuss the number of orientifolds. We determine the needed harmonic forms and related flux components. We then test in section III the solution 14 upon the different constraints \eqref{requir}. As announced, it does not succeed in satisfying all of them, but we give explicitly various values obtained, allowing to evaluate how far the solution is from a classical string background. The same procedure is followed in the appendix for solution 15, for which the results are worse. We finally relate the problem of classical de Sitter solutions to that of scale separation in section IV. We argue there that one should at best expect a bounded region in parameter space for both problems. We end with an outlook in section V.

\section{II. 6d geometry of solution 14}

In this section, we present in detail the geometry of the 6d group manifold for our de Sitter solution 14. We also obtain the material needed to test, in the next section, the requirements \eqref{requir} that would allow this solution to be a classical string background.

\subsection{A. Foreword on the change of basis}

As explained in the introduction, solutions 14 and 15 of \cite{Andriot:2020wpp} were found in a basis of one-forms associated to a 6d metric $\delta_{ab}$, $a=1,...,6$, giving respectively 8 or 7 non-zero structure constants. As detailed in section 2.3 and appendix C of \cite{Andriot:2020wpp}, a change of basis can be performed to new one-forms $e^a= e^a{}_m \d y^m$, associated to a new ``metric'' denoted $g_{ab}$
\beq
\d s_6^2 = g_{mn} \d y^m \d y^n = g_{ab} e^a e^b \ .\label{6dmetric}
\eeq
The coordinates $y^{m \,= \, 1, \, \dots, \, 6}$ parameterize circles: $y^m \in [0 , 2\pi[$ and we require the identifications $y^m \sim y^m + 2 \pi$. As we will see, the radii $r^{a} > 0$ are inside the $e^a{}_m$. The metric $g_{ab}$ has the interesting properties of being block diagonal along the pairs of $a$-indices $(12)$, $(34)$, $(56)$, along which are the sources, and the determinant of these blocks is equal to 1. For both solutions, the new basis gives only 4 structure constants. The corresponding algebras are then easy to identify: solution 14 is on $\mathfrak{g}_{3.5}^{0} \oplus \mathfrak{g}_{3.5}^{0}$, and solution 15 is on $\mathfrak{g}_{3.4}^{-1} \oplus \mathfrak{g}_{3.4}^{-1}$, using notations of \cite{Bock}. Both admit lattices, allowing the group manifolds to be compact. We come back to those in detail below. The new one-forms were denoted with a prime in \cite{Andriot:2020wpp}, and the solution data (flux components, structure constants, etc.) was given explicitly in appendix A of \cite{Andriot:2020wpp} under the names solution 14${}'$ and 15${}'$. We refer to section 2.3 and appendices A and C for more details on these solutions; in this paper, we work in the new basis and drop the prime. We now focus on solution 14, while solution 15 is treated in the appendix.

\subsection{B. Lattice, orientifolds}

For solution 14, the 4 structure constants are given by
\bea
&f^2{}_{35}= -0.28930 \ ,\ f^3{}_{25} = 0.013433 \ ,\\
&f^1{}_{64} = - 0.67154 \ , \ f^6{}_{14} = 0.41310 \ .\nn
\eea
This corresponds to two copies of the three-dimensional solvable algebra $\mathfrak{g}^0_{3.5}$. We rewrite them in terms of real positive numbers $N_{1,2,3,6} >0$ and the radii $r^a$, as in the following Maurer-Cartan equations
\bea
& \!\!\!\!\! \d e^2 = \frac{N_2 r^2}{r^3 r^5} e^3 \w e^5 \ , \ \d e^3 = - \frac{N_3 r^3}{r^2 r^5} e^2 \w e^5 \ ,\ \d e^5=0 \\
& \!\!\!\!\! \d e^1 = \frac{N_1 r^1}{r^4 r^6} e^6 \w e^4 \ , \ \d e^6 = - \frac{N_6 r^6}{r^1 r^4} e^1 \w e^4 \ ,\ \d e^4=0 \ . \nn
\eea
Let us focus on one of the two copies. As discussed below, an expression for globally defined one-forms is given by
\bea
e^2 & = r^2 \left(\frac{N_2}{N_3}\right)^{\frac{1}{4}} \left( \cos(\sqrt{N_2 N_3} y^5) \d y^2 - \sin(\sqrt{N_2 N_3} y^5) \d y^3  \right) \nn\\
e^3 & = r^3 \left(\frac{N_3}{N_2}\right)^{\frac{1}{4}} \left( \sin(\sqrt{N_2 N_3} y^5) \d y^2 + \cos(\sqrt{N_2 N_3} y^5)  \d y^3  \right) \nn\\
e^5 & = r^5 \d y^5 \ .\label{eag3.5}
\eea
Their normalisation is such that $e^2 \w e^3 = r^2 r^3 \d y^2 \w \d y^3$. The forms $e^{2,3}$ can be written in terms of the rotation matrix
\beq
A(y^5) = \left(\begin{array}{cc} \cos(\sqrt{N_2 N_3} y^5) &  - \sin(\sqrt{N_2 N_3} y^5) \\ \sin(\sqrt{N_2 N_3} y^5) & \cos(\sqrt{N_2 N_3} y^5) \end{array} \right)  \ ,\label{rot}
\eeq
which plays an important role. One verifies that the one-forms $e^a$ are globally defined, meaning invariant under $y^5 \sim y^5 + 2 \pi$, thanks to a coordinate identification
\beq
\left( y \right)_{y^5 + 2 \pi} = A(-2 \pi)\, \left( y \right)_{y^5} \ ,
\eeq
where the 2-vector $\left( y \right)$ stands for $y^2,y^3$. Such an identification is admissible if the entries of $A(-2 \pi)$ are integer (more generally, one may also allow for shifts of coordinates by multiples of $2\pi$). This gives the lattice quantization conditions, and ensures at the same time that we have globally defined one-forms. We refer to \cite{Andriot:2010ju} for more details. So here, a first possibility is to have $\sqrt{N_2 N_3} \in \mathbb{N}^*$. In that case, the coordinates are simply mapped to themselves, i.e.~they are globally defined: one is back topologically to a torus $T^6$, with a non-Ricci flat metric. A second possibility is $\sqrt{N_2 N_3} \in \mathbb{N} + \frac{1}{2}$, where the rotation gluing acts as a $\mathbb{Z}_2$ on the coordinates. Another possibility is $\sqrt{N_2 N_3} \in \mathbb{N} + \frac{1}{4}$ that mixes $y^{2}$ and $y^3$. These different lattices give rise to different topologies. We refer to \cite{Grana:2013ila} or \cite{Andriot:2015sia} (sections 2.3 and 5) for more details on these geometries. Further discussions and Betti numbers can also be found in \cite{Grana:2013ila}, as well as two more lattices involving constant shifts of coordinates, not included here. Ricci flat versions were recently considered in \cite{Acharya:2019mcu, Acharya:2020hsc}.

In the case of a torus, with $\sqrt{N_2 N_3} \in \mathbb{N}^*$, each coordinate is that of a circle without further identifications. Our orientifold involution conditions can then be mapped to conditions on coordinates: $\sigma(e^2) = e^2, \sigma(e^{3,5}) = - e^{3,5}$ are equivalent to $\sigma(y^2) = y^2, \sigma(y^{3,5}) = - y^{3,5}$, as can be seen through the forms \eqref{eag3.5}. This involution action on coordinates is the standard one on circles, so the counting of fixed points is the usual one, i.e.~2 per transverse circle. This gives $N_{O_5}^I = 2^4 = 16$. In the following, having this lattice in mind, we will impose the bound $N_s^I \leq 16$. For the other, more complicated lattices, this bound could be lowered, but smaller values for $N_s^I$ will as well be obtained.

\subsection{C. Harmonic forms}

With the above basis $\{e^a\}$ of globally defined one-forms, we can now determine the harmonic 1- and 3-forms with constant coefficients \cite{foot1}: those will be needed for flux quantization. To that end, we look for all closed and co-closed 1- and 3-forms \cite{foot2}. This is complicated due to the Hodge star, involving here the inverse metric \eqref{6dmetric} which has off-diagonal components $g^{ab}$. As a warm-up, let us consider the metric to be $\delta_{ab}$. In that case, one obtains $e^4, e^5$ and $e^1 \w e^4 \w e^6,\ e^1 \w e^5 \w e^6,\ e^2 \w e^3 \w e^4,\ e^2 \w e^3 \w e^5$. These forms are representative of cohomology equivalence classes. Changing the metric to $g_{ab}$ should not change these classes, since the latter are topological, so we should get the same number of forms. In addition, the representatives may only differ by an exact piece. This is what we obtain by an explicit computation: the 1-forms remain $e^4 , e^5$, while the harmonic 3-forms with constant coefficients are now
\bea
& \omega_1 = e^1 \w e^4 \w e^6 + \d o_1 \ ,\ \omega_2 = e^2 \w e^3 \w e^5 + \d o_2 \label{omegas}\\
& \omega_3 = e^1 \w e^5 \w e^6 + \d o_3 \ ,\ \omega_4 = e^2 \w e^3 \w e^4 + \d o_4 \ ,\nn
\eea
with the following exact pieces
\bea
o_1 & = \alpha \f{1}{46} \Big(  e^1 \w e^2 + \frac{\f{1}{46}}{\f{3}{25}} g^{56} e^3 \w e^4 \Big) \label{exactomegas} \\
& - \frac{\alpha \f{1}{46} \f{2}{35}  g^{34}+ g^{56}}{\f{6}{14}} e^5 \w e^6 \nn\\
o_2 & = \alpha \f{2}{35} \Big( - e^1 \w e^2  + \frac{\f{2}{35}}{\f{6}{14}} g^{34} e^5 \w e^6 \Big) \nn\\
& - \frac{\alpha \f{2}{35}\f{1}{46}g^{56}+ g^{34}}{\f{3}{25}}  e^3 \w e^4 \nn\\
o_3 & = \frac{ g^{12} \f{3}{25} }{(\f{6}{14})^2 + (\f{3}{25})^2 (1 + (g^{12})^2)}   e^3 \w e^6 \nn\\
o_4 & = \frac{ g^{12} \f{6}{14} }{(\f{6}{14})^2 + (\f{3}{25})^2 (1 + (g^{12})^2)}   e^3 \w e^6 \nn\\
& \!\!\!\!\! {\rm where}\ \  \alpha= \frac{ g^{12}  }{(\f{2}{35})^2 + (\f{1}{46})^2 (1 + (g^{12})^2)} \ .\nn
\eea
We verify that those are non-zero only because of the (inverse) metric off-diagonal components.

To perform the quantization, we will need to normalise these harmonic forms. Using the duality to the homology, and changing representatives of a same (co)homology class, we have the following equalities
\bea
\int_{\Sigma_3} \!\! \omega_1 = \int_{\tilde{\Sigma}_3} \!\! e^1 \w e^4 \w e^6 & = r^1 r^4 r^6 \int \! \int \! \int_{0}^{2\pi} \!\! \d y^1 \w \d y^4 \w \d y^6 \nn\\
& = (2\pi)^3 r^1 r^4 r^6 \ ,\label{omeganorm}
\eea
and the same holds for all four $\omega_i$. Indeed, the $e^a \w e^b \w e^c$ involved are precisely those that give $\d y^m \w \d y^n \w \d y^p$ times an appropriate (constant) normalisation. This is consistent with the formulas used so far for flux quantization, given that the only flux components that need to be quantized are those along harmonic forms. We have just verified here that these forms can be integrated, and the result is in agreement with the normalisation ansatz used so far, that relates the radii $r^a$ to the integral of $e^a$. The formula \eqref{qtties} for the quantized flux can thus be consistently used on the harmonic components.

\subsection{D. Flux quantization}

Fluxes should be quantized if and only if they are harmonic forms. This condition comes from having a flux $F = \d A$, where $A$ is locally but not globally defined, i.e.~$F$ is closed but not exact, in other words harmonic. It is then the transition function or patching of the gauge potential $A$ that gives the quantization condition. Let us consider the equations solved by the fluxes in our solutions
\bea
& \d F_1 = 0 \ , \ \d *_6 F_1 = 0 \ ,\quad \d F_3 \neq 0 \ ,\ \d *_6 F_3 = 0 \ ,\label{eqflux}\\
& \d H = 0 \ , \ \d *_6 H \neq 0 \ . \nn
\eea
$F_1$ is thus harmonic, while $F_3$ and $H$ are not but may contain pieces which are. Only the harmonic parts of these fluxes should be quantized. Doing such a proper flux quantization requires a good knowledge of the geometry, and in particular of the harmonic forms. The lack of such a knowledge for most solutions of \cite{Andriot:2020wpp} led us to quantize all flux components, as also done in the literature, but this is overconstraining. Here, using the harmonic 1- and 3-forms identified above for solution 14, we are able to express our fluxes within the general Hodge decomposition
\beq
F_q = F_{q\, {\rm harmo}} + \d A + *_6 \d B \ ,
\eeq
where $F_{q\, {\rm harmo}}, A, B$ are globally defined forms. We obtain more explicitly for solution 14
\bea
& H=\d\left( b_{13}\, e^1\w e^3 + b_{24}\, e^2\w e^4 \right) \ , \ F_1 = F_{1\, 5}\, e^5 \ , \label{fluxdecompo}\\
& F_3 = F_{3\, \omega_1}\, \omega_1 + F_{3\, \omega_2}\, \omega_2 \nn\\
& \phantom{F_3} + g_s^{-1} *_6 \d  \left( a_{12}\, e^1 \w e^2 + a_{34}\, e^3 \w e^4 + a_{56}\, e^5 \w e^6 \right) \ , \nn
\eea
where $g_s$ is introduced for notation convenience, and
\bea
& b_{13} = -0.34083 \ , \ b_{24}= 0.99383 \ , \ g_s F_{1\, 5}= -0.27398	 \ ,\nn\\
& g_s F_{3\, \omega_1} = 0.12430 \ ,\ g_s F_{3\, \omega_2} = 0.012539 \ ,\\
& a_{12} = 0.82025 \ ,\ a_{34} = -2.0877 \ , \ a_{56} = -0.55448\ .\nn
\eea
We deduce that only three flux components need to be quantized: $F_{1\, 5}, F_{3\, \omega_1}, F_{3\, \omega_2}$. To that end, we use the normalisation \eqref{omeganorm} for the harmonic forms, which justifies the use of the initial formulas \eqref{qtties}.

\section{III. Checking requirements for a classical solution}

With all the material obtained in the previous section, we are now ready to test solution 14 against the requirements \eqref{requir}. Given the supergravity solution data, namely the left-hand sides of \eqref{qtties}, we need to find 8 real parameters ($g_s, r^a, \lambda$) and 8 integers ($N_{1\, 5}, N_{3\, \omega_1}, N_{3\, \omega_2}, N_s^I, N_2 N_3 , N_1 N_6$), that satisfy the constraints \eqref{requir}. The source contributions are given by
\beq
\!\!\!\!\! g_s T_{10}^1 = 10,\ g_s T_{10}^2= -0.088507,\ g_s T_{10}^3 = -0.77652 \ .
\eeq
Solution 14 has the particularity of having $T_{10}^2 < 0$: this implies that among the $N_s^I$, an upper bound should only be imposed for $I=1$. We recall from \cite{Andriot:2020wpp} that $O_5$ may still be present in the set $I=2$.

For a better comparison to the results of \cite{Andriot:2020wpp}, we start by testing our solution without imposing the lattice conditions. We then get the following solution to the other constraints
\bea
& \mbox{No lattice condition}\!: \label{testnolatt}\\
& r^1= 86.658\ ,\ r^2= 272.28\ ,\ r^3= 10.834\ ,\ r^4= 18.142 \ ,\nn\\
& r^5= 198.25\ ,\ r^6= 10.562\ ,\ \lambda= 789.30\ ,\ g_s= 0.068818\ ,\nn\\
& N_{1\, 5}= -1  \ ,\ N_{3\, \omega_1} = 38 \ ,\ N_{3\, \omega_2} = 135 \ ,\nn\\
& N_s^1= 16 \ ,\ N_s^2= -17  \ ,\ N_s^3= -14  \ ,\nn\\
& N_2 = 0.0028913 \ ,\ N_3 = 0.084801 = (0.015659)^2/N_2 \ ,\nn\\
& N_1 = 0.0018814\ ,\ N_6= 0.077905 = (0.012107)^2/N_1\ ,\nn
\eea
where we give the highest products $N_1 N_6$ and $N_2 N_3$ obtained. Those remain far from $1^2$, but are less far from the other lattice allowing for $(1/4)^2$. We also obtain a solution to these constraints with $N_s^1= 14$, $N_2 N_3=1$ and $N_1 N_6 \approx 10^{-8}$. Solutions with lower $N_s^1$, down to $N_s^1 = 1$ can also be found. On the contrary, imposing lattice conditions (and the other requirements) without the bound on $N_s^1$ leads us at best to $N_s^1= 50960$, namely
\bea
& \mbox{No orientifold bound}\!:\\
& r^1= 57.907\ ,\ r^2= 162.65\ ,\ r^3= 10.014\ ,\ r^4= 354.99 \ ,\nn\\
& r^5= 2999.3\ ,\ r^6= 10.014\ ,\ \lambda= 186.97 \ ,\ g_s= 0.099885\ ,\nn\\
& N_{1\, 5}= -44 \ ,\ N_{3\, \omega_1} = 1370 \ ,\ N_{3\, \omega_2} = 3280 \ ,\nn\\
& N_s^1= 50960 \ ,\ N_s^2= -1195  \ ,\ N_s^3= -1241  \ ,\nn\\
& N_2 = 0.28572 \ ,\ N_3 = 3.5000 = 1^2/N_2 \ ,\nn\\
& N_1 = 0.22048 \ ,\ N_6= 4.5356 = 1^2/N_1\ .\nn
\eea
As expected, we can lower this $N_s^1$ when rather verifying the other lattice quantization conditions
\bea
& N_2 = 0.20794 \ ,\ N_3 = 1.2023 = \left(\tfrac{1}{2}\right)^2/N_2 \ ,\\
& N_1 = 0.10602\ ,\ N_6= 2.3581 = \left(\tfrac{1}{2}\right)^2/N_1\ ,\nn\\
\mbox{and}\ & N_s^1 = 13000 \ ,\nn\\
& N_2 = 0.56756 \ ,\ N_3 = 0.11012 = \left(\tfrac{1}{4}\right)^2/N_2 \ ,\\
& N_1 = 0.18581 \ ,\ N_6= 0.33636 = \left(\tfrac{1}{4}\right)^2/N_1\ ,\nn\\
\mbox{and}\ & N_s^1 = 3219\ ,\nn
\eea
and the other quantities verifying their constraints. These values of $N_s^1$ remain far too high.

Lattice conditions and the bound on $N_s^I$ have however to be respected in any case for the compactification to make sense, and we should rather test the string regime through the values of $g_s$ and the radii. Imposing all requirements but those on the radii, we find the solution
\bea
& \mbox{No large radius condition}\!:\\
& r^1= 7.4234\ ,\ r^2= 4.0198\ ,\ r^3= 0.24159\ ,\ r^4= 1.9131 \ ,\nn\\
& r^5= 16.164\ ,\ r^6= 0.11150\ ,\ \lambda= 1.0076\, ,\, g_s= 0.097666 \ ,\nn\\
& N_{1\, 5}= -45 \ ,\ N_{3\, \omega_1} = 2 \ ,\ N_{3\, \omega_2} = 2 \ ,\nn\\
& N_s^1= 14 \ ,\ N_s^2= -8  \ ,\ N_s^3= -18  \ ,\nn\\
& N_2 = 0.27890 \ ,\ N_3 = 3.5855 = 1^2/N_2 \ ,\nn\\
& N_1 = 0.019150\ ,\ N_6= 52.218 = 1^2/N_1\ .\nn
\eea
Two radii are substringy, i.e.~smaller than 1. We can also bring all of them to be greater than 1, except $r^6$ which gets lowered to $0.025984$. We get as well $r^{1,2,4,5}>10$ at the cost of having $r^3= 0.032184$ and $r^6 = 0.93577$. Using the lattice with $(1/4)^2$ does not change this situation. If rather we require large radii but relax the bound on $g_s$, we do not find any solution to the constraints \eqref{requir}. Note that a similar analysis and result has been obtained for a known de Sitter solution of type IIA supergravity in section 6 of \cite{Danielsson:2011au}. As there, we conclude that our de Sitter supergravity solution 14 cannot be made a classical string background.

\section{IV. Scale separation for de Sitter}

In this section, we discuss the matter of scale separation, and its relation to the requirements for a string classical regime. Consider a 10d (anti-)de Sitter solution with 4d cosmological constant $\Lambda$, and a tower of Kaluza--Klein states of mass scale $m_{KK}$ coming from the compactification on the 6d manifold. Having a 4d low energy effective theory, involving a finite number of degrees of freedom, is only possible if one truncates the tower of states by the 4d energy cutoff. This requires what is called scale separation, namely
\beq
\sqrt{|\Lambda|} \ll m_{KK} \ .
\eeq
It means that the typical 4d energy scale can be decoupled from the 6d one. In view of phenomenology, it is important to know whether scale separation can be achieved.

In many anti-de Sitter solutions, scale separation can actually not be reached: see e.g.~\cite{Blumenhagen:2019vgj, Apruzzi:2019ecr, Lust:2020npd, Farakos:2020phe, Emelin:2020buq} for recent papers on this point, and \cite{Caviezel:2008ik, Tsimpis:2012tu, Petrini:2013ika, Gautason:2015tig} and references therein for older related works. This has led to recent swampland conjectures \cite{Gautason:2018gln, Lust:2019zwm} forbidding scale separation (see also \cite{Font:2019uva, Buratti:2020kda, Andriot:2020lea}). Interestingly, one classical counter-example exists, the so-called DGKT anti-de Sitter solution \cite{DeWolfe:2005uu} (extended recently in \cite{Marchesano:2019hfb, Junghans:2020acz, Marchesano:2020qvg}). A criticism of this solution is the problem of ``smeared'' sources, and we refer to \cite{Andriot:2020wpp} for more references and discussion on this point. This example remains interesting here for two reasons. First, its framework is very analogous to ours: it is a classical 10d supergravity solution, on a torus, and its intersecting sources configuration is analogous to ours. Another similarity is the presence of flux integers not constrained through the tadpole or Bianchi identity: here these are the harmonic components of $F_3$, giving the $N_{3\, \omega_i}$. Secondly, the DGKT solution admits a parametric control (with asymptotic limit) on the scale separation, which is the same that governs the validity of the solution as a classical string background: the more classical (small $g_s$, large volume, etc.), the better the scale separation. This provides an example of a relation between these two important problems.

For de Sitter solutions, a similar relation was sketched in \cite{Andriot:2019wrs}. First, it was first shown there that a classical de Sitter solution requires a small 6d curvature scale $|{\cal R}_6|$ compared to that of the average internal length $2\pi r$: $|{\cal R}_6| \times (2\pi r)^2 \ll 1$. This is a priori possible on some manifolds thanks to internal hierarchies and/or fine-tunings. In addition, such a hierarchy in a de Sitter solution was shown to automatically imply scale separation, in the sense of
\beq
{\cal R}_4 \ll \frac{1}{(2\pi r)^2} \ .\label{preres}
\eeq
Let us test here scale separation with the de Sitter solution 14 in the case where we ignore lattice conditions \eqref{testnolatt}: all other requirements for a classical solution are then satisfied. With these values and the $\lambda$-rescaling, we first compute (in units of $2\pi l_s$)
\beq
(\Rc_4 = 3.6370 \cdot 10^{-8} ) < ( 1/(2\pi r)^2 = 1.1873 \cdot 10^{-5} ) \ , \label{scalesepex}
\eeq
where $r=(\prod r^a)^{\frac{1}{6}}$ \cite{foot4}. The scale separation condition \eqref{preres} is verified, since we consider here hierarchies of order $10$ per length. This result is however only preliminary, since the largest radius $r^2$ gives a smaller value, as well as the 6d curvature
\beq
\!\!\! 1/(2\pi r^2)^2 = 3.4167 \cdot 10^{-7}  \ ,\  |\Rc_6| = 1.2162 \cdot 10^{-6}  \ .
\eeq
A tower of modes associated to the radius $r^2$ would then be difficult to separate or truncate \cite{foot5}. The internal hierarchy $|{\cal R}_6| \times (2\pi r)^2  \ll 1$ is also not verified, due to the unsatisfied lattice conditions. The lack of complete classicality (in the sense of satisfying requirements \eqref{requir}) seems here again related to difficulties in having scale separation.

Even though classical de Sitter solutions may admit scale separation, we do not expect the same parametric control as in DGKT. As expressed in \cite{Andriot:2020wpp}, classical de Sitter solutions may not exist at parametric control in an asymptotic limit, but they could still be present in a bounded region of parameter space. In such a ``grey zone'', parameters are large/small enough to accommodate a classical regime as in the requirements \eqref{requir}, but they remain bounded by these same constraints and cannot be taken asymptotically. At best, one may then have scale separation in this bounded region, controlled by a bounded parameter. In the following, and later in \cite{footPaul}, we illustrate this idea with a simple transformation that allows us to move in parameter space. We use a scaling parameter $\gamma >1$ to act as follows on the entries in the right-hand sides of \eqref{qtties}
\beq
\!\!\!\! r^a \rightarrow \gamma^{x_a}\, r^a \ ,\ g_s \rightarrow \gamma^g\, g_s \ ,\ N_{K} \rightarrow \gamma^{n_K}\, N_K \ , \ \lambda \rightarrow \lambda  \label{g1}
\eeq
and we do not consider any quantized $H$-flux nor $F_5$ above because there are none in solutions 14 and 15. The powers $x_a, g, n_K$ and their signs are not fixed. However, we relate them in such a way that the left-hand sides of \eqref{qtties} remain invariant. Taking for simplicity $\forall a,\, x_a=x$, we get
\beq
n_{q=1,3} = qx-g \ ,\ \ n_{s}^I = 4x -g \ ,\ \ n_a = x \ .\label{g2}
\eeq
By not affecting the quantities in the left-hand sides of \eqref{qtties}, we are guaranteed to still have a solution to the 10d equations. ${\cal R}_4$ is also invariant under this simple transformation. The 6d radii are however changed, so the ratio of the cosmological constant to a Kaluza--Klein mass goes as the latter, and the scale separation is improved with smaller $r^a$, i.e.~$x<0$. This specific $\gamma$-scaling then does not help in getting classical solutions, which rather require large radii. Such a situation where two requirements go in different parametric directions is common for classical de Sitter solutions, due to the many bounds to satisfy. This leads to the bounded region in parameter space, as we illustrate in table \ref{tab:scaling} with the different possible signs of $x,g$, corresponding to various parametric directions.

We see from table \ref{tab:scaling} that in any direction of the $\gamma$-scaling, a parameter bound is met. Some cases in this table were also observed in the explicit example in section III. The simple $\gamma$-scaling was considered here for the purpose of illustration. Other transformations may do better on some of the requirements, e.g.~for the lattice conditions or the scale separation: we introduce one at the end of this paper in \cite{footPaul}, the $\beta$-scaling. Specificities of given solutions may also help. Nevertheless, it remains unlikely that an asymptotic limit would be opened in some parametric direction.

Let us also stress two differences with an anti-de Sitter solution, that are made more obvious with table \ref{tab:scaling}. An anti-de Sitter solution would allow for all $N_s^I \leq 0$, in which case there is no upper bound on $|N_s^I|$, contrary to the above $N_s^1 \leq N_{O_p}^1$. In addition, in the DGKT anti-de Sitter solution, one has $f^a{}_{bc}=0$, which is not possible for de Sitter solutions on group manifolds that require ${\cal R}_6 <0$. The former thus does not face the lattice conditions. Removing these two constraints relieves bounds in parameter space.

\begin{table}[H]
  \begin{center}
    \begin{tabular}{|c||c|c|}
    \hline
 & $x > 0$ & $x<0$  \\
    \hhline{=::==}
 &  & With $4x=g\, $:\\
 & classical \cgr{$\checkmark$}, & classical \cre{$\times$} ($r^a$ bound), \\
 $g<0$ & scale sep.~\cre{$\times$}, & scale sep.~\cgr{$\checkmark$},\\
 & fluxes, lattice \cgr{$\checkmark$}, & fluxes, sources \cgr{$\checkmark$}, \\
 & sources \cre{$\times$}  ($N_s^1 \leq N_{O_p}^1$)  & lattice \cre{$\times$} (integer cond.) \\
     \hhline{-||--}
 & With $4x=g\, $: &  \\
 & classical \cre{$\times$} ($g_s$ bound), & classical \cre{$\times$} ($r^a, g_s$ bounds), \\
 $g>0$ & scale sep.~\cre{$\times$}, & scale sep.~\cgr{$\checkmark$}, \\
  & lattice, sources \cgr{$\checkmark$}, & fluxes, sources, lattice \cre{$\times$} \\
  & fluxes \cre{$\times$} (integer cond.) & (integer cond.) \\
    \hline
    \end{tabular}
    \end{center}
     \caption{Starting with a de Sitter solution, we act with the $\gamma$-scaling \eqref{g1}, \eqref{g2} on the parameters. We indicate with a \cgr{$\checkmark$} or \cre{$\times$} whether or not it helps in meeting the various requirements. When it does not, the limiting bound is written in parentheses. The ``integer condition'' bound refers to integer numbers having to be greater than $1$. Since $N_s^1$ is very constrained, $1 \leq N_s^1 \leq N_{O_p}^1$ in a de Sitter solution, we restrict when possible to the scaling leaving this number invariant, namely $4x = g$.}\label{tab:scaling}
\end{table}

To conclude, if a region of classical de Sitter solutions exists, we do not exclude that scale separation also takes place there. We believe however that it would be at best controlled by a bounded parameter, contrary to the DGKT solution and despite several similarities with that framework. Such a bounded region or grey zone with classical and scale separated de Sitter solutions remains to be found \cite{foot6}, and we hope to come back to these questions in future work.

\section{V. Outlook}

In this work we have tested whether two 10d supergravity de Sitter solutions of \cite{Andriot:2020wpp} could be promoted to classical string backgrounds. Having identified the requirements \eqref{requir} to be met, we have derived all the necessary material to perform these checks, to a rare extent in the literature. Eventually, our solutions do not pass these tests, as they fail to satisfy simultaneously all requirements, despite partial positive results. We also discussed the relation of this problem to that of scale separation, and what one should expect on the latter for classical de Sitter solutions. We believe that the tools developed here and the explicit results obtained should be useful in future studies.

Whether supergravity solutions could be classical string backgrounds has already been partially analysed on de Sitter solutions of \cite{Andriot:2020wpp}. Solution 14 did not appear as the most favored one: four other solutions were doing better regarding the partial requirements imposed there. But for those, a complete knowledge of the 6d geometry was missing, preventing us to go further (see however the appendix here). It would be interesting to focus more on them, or more generally, to try with other solutions, possibly obtained from ours by small deformations or transformations, that could help in satisfying the requirements. A family of de Sitter solutions obtained by varying one parameter is depicted in figure 1 of \cite{Danielsson:2010bc}. In such a family, one may eventually find a bounded region of classical solutions, as discussed in section IV.

As it appears from this work, the question of whether one can obtain classical de Sitter solutions of string theory is not settled. Even though we have not found an example for now, our study highlighted how close one can get, and how involved these tests actually are. A related matter is the definition of the requirements for a classical solution: one may consider our requirements on $g_s$ and the radii as conservative. Can one allow for slightly higher values of $g_s$? What length or volume should actually be bigger than $l_s$ on our group manifolds? Precise answers to these questions depend on the corrections to effective theories, and those are typically difficult to determine, especially away from a more standard Ricci flat compactification. If any room could be gained from this side, it would certainly be interesting. We hope to come back to all these important problems in future work.

\section{Acknowledgements}

D.~A.~and P.~M.~acknowledge support from the Austrian Science Fund (FWF): project number M2247-N27. P.~M.~thanks the ITP at TU Wien for hospitality and for the opportunity to work on this project. T.~W.~acknowledges support from the Austrian Science Fund (FWF): project number P 30265.

\appendix*
\setcounter{equation}{0}
\section{Appendix: Solution 15}

The 4 structure constants of solution 15 are given by
\bea
& f^2{}_{35} = -0.60208 \ ,\ f^3{}_{25} = -0.058853 \ ,\\
& f^4{}_{61} = -0.10206 \ ,\ f^6{}_{41} = -0.015345 \ ,\nn
\eea
giving two copies of the solvable algebra $\mathfrak{g}^{-1}_{3.4}$. We focus here on one. We rewrite it as
\beq
\d e^2 = \frac{N_2 r^2}{r^3 r^5} e^3 \w e^5 \ , \ \d e^3 = \frac{N_3 r^3}{r^2 r^5} e^2 \w e^5 \ ,\ \d e^5=0 \ ,
\eeq
with real positive numbers $N_{2,3} >0$. Globally defined one-forms can be written as in \eqref{eag3.5} where one replaces the rotation matrix \eqref{rot} by the following ``weighted hyperbolic rotation matrix''
\beq
\left(\begin{array}{cc} \! \cosh(\sqrt{N_2 N_3} y^5) & \!\!\!\! -\left(\frac{N_3}{N_2}\right)^{\frac{1}{2}} \sinh(\sqrt{N_2 N_3} y^5) \\ \! -\left(\frac{N_2}{N_3}\right)^{\frac{1}{2}} \sinh(\sqrt{N_2 N_3} y^5) & \!\!\!\! \cosh(\sqrt{N_2 N_3} y^5) \end{array} \right) \nn
\eeq
This gives the normalisation $e^2 \w e^3 = r^2 r^3 \d y^2 \w \d y^3$ \cite{foot3}. The above matrix evaluated at $-2\pi$ provides again the lattice quantization conditions, by requiring the entries to be integers. We obtain \cite{Andriot:2010ju}
\beq
\!\!\!\! \cosh(\sqrt{N_2 N_3} 2\pi) = n_1,\ n_1^2 - n_2n_3=1, \ \frac{N_3}{N_2}=\frac{n_3}{n_2} \ ,
\eeq
for $n_{1,2,3} \in \mathbb{N}^*$. Note that $n_2=n_3$ is not a solution to these conditions, hence the need of the ``weight'' in the hyperbolic rotation.

The harmonic 3-forms with constant coefficients in $\{e^a\}$ basis are formally the same as \eqref{omegas} and \eqref{exactomegas}, up to the relabeling $1\leftrightarrow 4$, $2\leftrightarrow 3$, and setting $g^{56}=0$; this is how we define $\omega_{1,2}$ here below. This can be understood by looking at the algebras and the metrics. Similarly, the harmonic 1-forms are now $e^1, e^5$. We then obtain for this solution the following Hodge decomposition of the fluxes
\bea
& H=\d\left(  b_{24}\, e^2\w e^4 \right) \ , \ F_1 = F_{1\, 5}\, e^5 \ , \label{fluxdecompo2}\\
& F_3 = F_{3\, \omega_1}\, \omega_1 + F_{3\, \omega_2}\, \omega_2 \nn\\
& \phantom{F_3}  + g_s^{-1} *_6 \d  \left( a_{12}\, e^1\w e^2 + a_{34}\, e^3\w e^4 + a_{56}\, e^5\w e^6 \right) \ , \nn
\eea
where
\bea
& b_{24}= 0.0018085 \ , \ g_s F_{1\, 5}= 0.13944 \ ,\\
& g_s F_{3\, \omega_1} = 0.0014004 \ ,\ g_s F_{3\, \omega_2} = 0.00013677 \ , \nn\\
\ & a_{12} = 1.0117 \ ,\ a_{34} = 0.048104 \ , \ a_{56} = -0.15985\ . \nn
\eea
This gives again only three flux components to quantize. For completeness, we finally give the source contributions
\beq
\!\!\! g_s T_{10}^1 = 10,\ g_s T_{10}^2= 0.49663,\ g_s T_{10}^3 = -0.10585 \ .
\eeq

We now test this solution as done above for solution 14. Without imposing the lattice conditions we obtain a first solution to the other requirements \eqref{requir}
\bea
& \mbox{No lattice condition}\!: \label{testnolatt15}\\
& r^1= 108.00\ ,\ r^2= 47.861\ ,\ r^3= 21.302\ ,\ r^4= 12.912 \ ,\nn\\
& r^5= 25.270\ ,\ r^6= 10.826\ ,\ \lambda= 2372.7\ ,\ g_s= 0.0014851\ ,\nn\\
& N_{1\, 5}= 1 \ ,\ N_{3\, \omega_1} = 6 \ ,\ N_{3\, \omega_2} = 1 \ ,\nn\\
& N_s^1= 15\ ,\ N_s^2= 14 \ ,\ N_s^3= - 3 \ ,\nn\\
& N_2 = 0.0028540 \ ,\ N_3 = 0.0014083 \ ,\nn\\
& N_4 = 0.0038956\ ,\ N_6= 0.00083305 \ .\nn
\eea
We find other solutions down to $N_s^{1}=N_s^{2}=1$. On the contrary, imposing the lattice quantization conditions is more difficult. Imposing no bound on the number of sources $N_s^{1,2}$, we can satisfy the other constraints if we trade the integer conditions for numbers greater than $1$. We then get a solution with $N_s^{1}\approx 755.40$, $N_s^{2}\approx 1746.1$, which remains too high. Finally, imposing only the lattice conditions, the bound on $N_s^I$ and the flux quantization, to test the string regime on $g_s$ and $r^a$, we could not find any solution. These results are worse than for solution 14. The de Sitter supergravity solution 15 cannot be made a classical string background.

Let us finally comment on solutions 10 and 12 of \cite{Andriot:2020wpp}: those were found in that paper to satisfy partial requirements for classical string backgrounds. The algebras of these two solutions were not identified in \cite{Andriot:2020wpp}, but we achieve here this identification. Indeed, we have found a change of basis transforming their structure constants to the following sets of non-zero $f^a{}_{bc}$
\bea
\mbox{Solution 10:}&\quad  f^4{}_{26} = f^6{}_{42} = - f^2{}_{64} = 1 \ ,\\
&\quad f^3{}_{15} =  f^1{}_{35} = - 1 \ ,\nn\\
\mbox{Solution 12:}&\quad  f^4{}_{61} = f^6{}_{14} = - f^1{}_{46} = 1 \ ,\\
&\quad f^3{}_{25} = - f^2{}_{35} = 1 \ ,\nn
\eea
allowing to identify the corresponding algebras: $\mathfrak{s}\mathfrak{o}(2,1) \oplus \mathfrak{g}^{-1}_{3.4}$ for solution 10, and $\mathfrak{s}\mathfrak{o}(2,1) \oplus \mathfrak{g}^{0}_{3.5}$ for solution 12. The $\mathfrak{s}\mathfrak{o}(2,1)$ part leads to a non-compact group manifold, so these solutions are actually not valid as compactifications. This is reminiscent of the analysis of \cite{Dibitetto:2010rg} and references therein, where de Sitter solutions can be found in 4d gauged supergravities with semi-simple gaugings corresponding to non-compact spaces.


\begin{thebibliography}{100}
\bibitem{foot1}
Further harmonic forms can be obtained in the $\{e^a\}$ basis with globally defined functions as coefficients, such as $\cos(\sqrt{N_2 N_3} y^5)$, etc.; we will not need those.
\bibitem{foot3}
Slightly different expressions of the one-forms would allow for the same normalisation. In particular another one more analogous to \eqref{eag3.5} is possible, but it would not lead to an appropriate lattice quantization.
\bibitem{foot4}
For completness and illustration, we compute for this set of values the square of the Planck mass in units of $2\pi l_s$
\bea
& (2\pi l_s)^2 \times  M_p^2 = \frac{(2\pi l_s)^2}{\kappa_{10}^2} \int_6 \d^6 y \sqrt{|g_6|}\, g_s^{-2} = \frac{4\pi}{ g_s^2} \prod_{a=1}^{6} \frac{r^a}{l_s} \nn \\
& = 2.5767 \cdot 10^{13} = ( 5.0761 \cdot 10^6 )^2 \ .\label{Mp}
\eea
\end{thebibliography}
\end{document}